\documentclass[aps,prx,twocolumn,colorlinks=true,linkcolor=black,urlcolor=blue,citecolor=black,pdfencoding=auto]{revtex4-1}
\usepackage{graphicx}
\usepackage{color}
\usepackage{amsmath}
\usepackage{amsfonts}
\usepackage{bbold}
\usepackage{amssymb}
\usepackage{subfigure}
\usepackage[normalem]{ulem}
\definecolor{diffcolor}{RGB}{175,31,36}
\usepackage{hyperref}

\begin{document} 
\title{Bounds on the superconducting transition temperature: Applications to twisted bilayer graphene and cold atoms} 

\author
{Tamaghna Hazra,$^{1}$ Nishchhal Verma,$^{1}$ Mohit Randeria$^{1\ast}$\\
\normalsize{$^{1}$Department of Physics, The Ohio State University, Columbus, Ohio 43210}
}
\date{\today}


\begin{abstract}
Understanding the material parameters that control the superconducting transition temperature $T_c$ 
is a problem of fundamental importance. In many novel superconductors phase fluctuations  
determine $T_c$, rather than the collapse of the pairing amplitude. 
We derive rigorous upper bounds on the superfluid phase stiffness for multi-band systems, valid in any dimension. 
This in turn leads to an upper bound on $T_c$ in two dimensions (2D), which holds irrespective of pairing mechanism,
interaction strength, or order-parameter symmetry. 
Our bound is particularly useful for the strongly correlated regime of low-density and narrow-band systems, 
where mean field theory fails. For a simple parabolic band in 2D with Fermi energy $E_F$, we find that 
$k_BT_c \leq E_F/8$, 
an exact result that has direct implications for the 2D BCS-BEC crossover in ultra-cold Fermi gases. 
Applying our multi-band bound to magic-angle twisted bilayer graphene (MA-TBG), we find that
band structure results constrain the maximum $T_c$ to be close to the experimentally observed value.
Finally, we discuss the question of deriving rigorous upper bounds on $T_c$ in 3D.
\end{abstract}

\maketitle 


Our work is motivated by the fundamental question: what limits the superconducting (SC) transition temperature $T_c$?
Within BCS mean-field theory, and its extensions like Eliashberg theory, the amplitude of the SC order parameter
is destroyed by the breaking of pairs, and $T_c$ scales with the pairing gap $\Delta$. The material parameters that control the
mean-field $T_c$ are the electronic density of states (DOS) at the chemical potential $N(0)$
and the effective interaction, determined by the spectrum of fluctuations that mediate pairing. 

Beginning with the pioneering experiments of Uemura~\cite{UemuraPhys.Rev.Lett.1989} and theoretical ideas of 
Emery and Kivelson~\cite{EmeryNature1995} on underdoped cuprates, it became clear that the mean field picture of $T_c$ 
scaling with the pairing gap is simply not valid in many novel superconductors.
The loss of SC order is then governed by fluctuations of the phase of the order parameter, rather than the suppression of 
its amplitude, and $T_c$ is related to the superfluid stiffness $D_s$. The material parameters that determine
$D_s$ are rather different from those that determine the pairing gap $\Delta$. 

The question of mean field amplitude collapse versus phase fluctuation dominated SC transition
is brought into sharp focus by a variety of recent experiments in narrow band and low density systems.
One of the most exciting recent developments
is the observation of very narrow bands in magic-angle twisted bilayer graphene (MA-TBG) leading to correlation-induced ``Mott'' 
insulating states~\cite{CaoNature2018a} and superconductivity~\cite{CaoNature2018} in their vicinity.
Flat bands are also also expected to arise in various topological states of matter;
see, e.g.,~\cite{KopninPhysRevB2011,TangNatPhys2014,PeottaNatComm2015,LiangPhysRevB2017}.
BCS theory-based intuition suggests that narrow bands have a large DOS $N(0)$ and
lead to high temperature superconductivity. Is this true or do phase fluctuations limit the $T_c$?

The extensive compilation of data in Fig.~6 of ref.~\cite{CaoNature2018} suggests that
all known superconductors have a $T_c$ that scales at most like a constant times 
the ``Fermi energy $E_F$", though there is considerable leeway in defining 
$E_F$ in strongly correlated and multi-band materials.
We also note that ultra-cold Fermi gases in the strongly interacting regime of the BCS-BEC 
crossover~\cite{ketterle.zweirlein.2006,RanderiaAnnu.Rev.Condens.MatterPhys.2014} 
exhibit experimental values~\cite{KuScience2012} of $k_B T_c/E_F$ larger than those observed in the solid state. 
All of these observations raise the question of ultimate limits on the $T_c$ of a superconductor or paired superfluid.

In this paper, we obtain sharp answers to these questions, especially in 2D.
First, we derive an upper bound on the superfluid stiffness $D_s(T) \leq \widetilde{D}(T)$,
where $\widetilde{D}$ is proportional to the optical conductivity sum rule.
This inequality is valid in all dimensions and for arbitrary interactions.
We then use the Berezinskii-Kosterlitz-Thouless (BKT) theory in 2D
to obtain $k_B T_c \leq \pi \widetilde{D}(T_c)/2$.

While the bound on $T_c$ is of completely general validity, it is most useful
in the strongly correlated regime of narrow-band and low density systems,
precisely where conventional mean-field approaches fails.
We show that $\widetilde{D}$ is necessarily ``small" in 
such systems, and, in many cases of interest,
$\widetilde{D}$ is essentially determined by the (non-interacting) band structure. 

We give several examples that illustrate the usefulness of our bounds for a variety of systems. 
For a single parabolic band we show that $k_B T_c \leq E_F/8$ in 2D.
This exact result poses stringent constraints on the $T_c$ of
the 2D BCS-BEC crossover in ultra cold atoms.
We also describe bounds on $T_c$ for the 2D attractive Hubbard model, relevant for current optical
lattice experiments~\cite{MitraNaturePhys2018}, that demonstrate the tension
between breaking of pairs and phase fluctuations, and highlight the connection 
with a pairing pseudogap~\cite{RanderiaPhys.Rev.Lett.1992,TrivediPhys.Rev.Lett.1995}.

Turning to multi-band systems, we use available band structure 
results~\cite{BistritzerProc.Natl.Acad.Sci.2011,LopesdosSantosPhys.Rev.Lett.2007,
KoshinoPhys.Rev.X2018a,KangPhys.Rev.X2018,PoPhys.Rev.X2018}
for MA-TBG to calculate $\widetilde{D}$ and thus
constrain its $T_c$ without any assumptions about the pairing mechanism or order-parameter symmetry. 
We obtain a rigorous (but weak) bound of $\simeq 15$ K.
Using physically motivated approximations, we estimate a bound on $T_c$ as low as 6 K.

Finally, we discuss the question of deriving similar bounds in 3D. We show that the presence of
non-universal pre-factors in the relation between $T_c$ and $D_s$, as well their scaling behavior near
a SC quantum critical point, pose challenges in deriving a rigorous bound in 3D.

{\bf Results:} We first outline our main results and then give a detailed derivation and specific applications. 
We consider a Fermi system described by the general Hamiltonian
\begin{equation}
    {\cal H} = {\cal H}_K +  {\cal H}_{\rm int}; \ \ \ {\cal H}_K = \sum_{{\bf k},m,\sigma}\epsilon_m({\bf k})c^\dag_{{\bf k}m\sigma}c^{\phantom{\dag}}_{{\bf k}m\sigma} \label{eq:H}
\end{equation}
where {\bf k} is crystal momentum, $m$ is a band label, and $\sigma$ the spin. 
${\cal H}_K$ is the kinetic energy and ${\cal H}_{\rm int}$ describes interactions (electron-phonon, electron-electron, etc.),
including those that give rise to superconductivity. The external vector potential ${\bf A}$ enters ${\cal H}$ through a 
Peierl's substitution in the tight-binding representation of ${\cal H}_K$, but does not affect ${\cal H}_{\rm int}$. 
For now, we ignore disorder and return to it at the end.

The macroscopic superfluid stiffness $D_s$ determines the free energy cost of distorting the 
phase of the SC order parameter $|\Delta|e^{i\theta}$ via the Boltzmann factor $\exp\left(- D_s \int d^d{\bf r}|\nabla\theta|^2|/2k_B T\right)$.
It is related to the London penetration depth via $1/\lambda_L^2 = (4\mu_0 e^2/\hbar^2)D_s$ in 3D. Microscopically, 
$D_s$ can be calculated as the static, long wavelength limit of the transverse current response~\cite{Baym.1968,ScalapinoPhys.Rev.B1993}
to a vector potential. 
(Our results are equally valid for neutral superfluids with rotation playing the role of the magnetic field.)
We obtain a rigorous upper bound valid in any dimension
\begin{equation}
    D_s(T) \ \le\ \widetilde{D}(T) = {{\hbar^2}\over{4\Omega}}\,\sum_{{\bf k},mm',\sigma}\,
    M^{-1}_{mm'}({\bf k})\,\langle c^\dag_{{\bf k}m\sigma}c^{\phantom\dag}_{{\bf k}m'\sigma} \rangle
\label{eq:ineq}
\end{equation}
where $\Omega$ is the volume of the system and $M^{-1}_{mm'}({\bf k})$ is an inverse mass tensor that depends only on
the electronic structure of ${\cal H}_K$; see eq.~(\ref{eq:mass}) below.
The temperature and interactions impact $\widetilde{D}$ only through 
$\langle c^\dag_{{\bf k}m\sigma}c^{\phantom\dag}_{{\bf k}m'\sigma} \rangle$,
where the thermal average is calculated using the full ${\cal H}$.

We next use $D_s$ to provide an upper bound on the SC transition temperature in 2D.
We use the Nelson-Kosterlitz~\cite{NelsonPhys.Rev.Lett.1977} universal relation to obtain
\begin{equation}
k_B T_c \le \pi \widetilde{D}(T_c)/2
\label{eq:2DTcbound}
\end{equation}
For a weak coupling superconductor, $T_c$ is well described by mean field theory and our result, though valid as an upper bound, may not be very useful.
On the other hand, as we show below, for a strongly interacting system 
the bound gives insight into both the value
of $T_c$ and on its dependence on parameters.

{\bf Bound on superfluid stiffness:} The intuitive idea behind $D_s \leq \widetilde{D}$ is as follows. 
$\left(2\pi e^2/\hbar^2\right)\widetilde{D} = \int_0^\infty d\omega\,{\rm Re}\,\sigma(\omega)$ is the
optical conductivity spectral weight integrated over the bands in eq.~(\ref{eq:H}), 
and $\left(4\pi e^2/\hbar^2\right)D_s$ is the coefficient of
the $\delta(\omega)$ piece in ${\rm Re}\,\sigma(\omega)$ in the SC state;
(note: $\int_0^\infty d\omega\,\delta(\omega)=1/2$).
The inequality~(\ref{eq:ineq}) says that the weight in the SC delta-function
must be less than or equal to the total spectral weight.

To derive (\ref{eq:ineq}), we use the Kubo formula for $D_s$ as 
a linear response~\cite{Baym.1968,ScalapinoPhys.Rev.B1993} to an external vector potential in an arbitrary direction $a$
\begin{equation}
  D_s = \widetilde{D} - \left({\hbar^2}/{4e^2}\right)\ \chi_{j_a j_a}^\perp( {\bf q }\rightarrow 0 ,\omega =0),
\label{eq:kubo1}
\end{equation}
where $\widetilde{D}$ is the diamagnetic response $\sim\!\left\langle\delta^2{\cal H}/\delta{A_a}^2\right\rangle$, while $\chi^\perp$ is
the {\it transverse} current-current correlation function. 
$\widetilde{D}$ is given by eq.~(\ref{eq:ineq}) with 
\begin{equation}
M^{-1}_{mm^\prime}({\bf k}) = \sum\limits_{\alpha\beta}\ 
U^\dag _{m,\alpha}({\bf k}) \ {{\partial^2 t_{\alpha\beta}({\bf k})}\over{\partial(\hbar k_a)^2}}\ U_{\beta,m^\prime}({\bf k})
\label{eq:mass}
\end{equation}
Here $\alpha,\beta$ label orbitals/sites within a unit cell of a Bravais lattice,
$t_{\alpha\beta}({\bf k})$ is the Fourier transform of the hopping $t_{\alpha\beta}({\bf r}_{i\alpha} - {\bf r}_{j\alpha})$,
and $U_{\alpha,m}({\bf k})$ is the unitary transformation that diagonalizes $t_{\alpha\beta}({\bf k})$ 
to the band basis $\epsilon_m({\bf k})\delta_{m,m'}$. The inverse mass tensor in eq.~(\ref{eq:mass}) also depends on the
direction $a = x,y, \ldots$ through the derivative with respect to $k_a$ on the right hand side, however, we
do not show this $a$ dependence explicitly to simplify the notation.
These results are derived in Appendix~\ref{sec:kubo}, and 
the relation to the optical sum rule shown in Appendix~\ref{sec:optical}; see also
ref.~\cite{ValenzuelaPhys.Rev.B2013}.

We next turn to the second term in eq.~(\ref{eq:kubo1}).
From its Lehmann representation 
we see that $\chi^\perp({\bf q}\!\rightarrow\!0,\omega\!=\!0) \geq 0$ at all temperatures;
see Appendix~\ref{sec:para}. We thus obtain $D_s(T) \le \widetilde{D}(T)$. 

For a single band system eqs.~(\ref{eq:ineq}) and (\ref{eq:mass}) simplify greatly and we get
$\widetilde{D} = (4\Omega)^{-1}\sum_{{\bf k},\sigma}(\partial^2 \epsilon({\bf k})/{\partial k_a^2})\ n_{\sigma}({\bf k})$,
where the momentum distribution $n_{\sigma}({\bf k}) = \langle c^\dag_{{\bf k}\sigma}c^{\phantom\dag}_{{\bf k}\sigma} \rangle$.
This allows us to recover well-known special cases.
(1) With nearest neighbor (NN) hopping on a square or cubic lattice,  $ {\partial^2 \epsilon({\bf k})}/{\partial k_a^2}\!\sim\!\epsilon({\bf k})$, and
$\widetilde{D}$ is proportional to the kinetic energy. 	
(2) A parabolic dispersion $\epsilon({\bf k}) = \hbar^2 k^2/2m$ leads to the simple result
$\widetilde{D} = \hbar^2 n/{4m}$, independent of $T$ and of interactions.
Here $D_s(T) = \hbar^2 n_s(T)/{4m}$ and our bound simply says that the superfluid density $n_s(T) \le n$ the total density.

For materials with non-parabolic dispersion and/or multiple bands,
$\widetilde{D}$ depends on $T$ and interactions.
It is thus illuminating to derive a bound for $\widetilde{D}$ which depends only on the density.
We describe the single band result here, relegating 
the multi-band generalization to Appendix~\ref{sec:realspace}.
We write $\mathcal{H}_K = - \sum_{{\bf R}{\boldsymbol{\delta}}\sigma} 
 \big[ t(\boldsymbol{\delta}) c^\dagger_{{\bf R} +\boldsymbol{\delta},\sigma} c^{\phantom\dagger}_{{\bf R},\sigma} + {\rm h.c.} \big]$
with translationally invariant hopping amplitudes $t(\boldsymbol{\delta})$ that depend only the vector $\boldsymbol{\delta}$ connecting lattice sites ${\bf R}$ and ${\bf R} +\boldsymbol{\delta}$.
We couple the system to a vector potential and compute $\widetilde{D}$, which involves
terms like
$\sum_{i,j} {\delta}_a^2 t(\boldsymbol{\delta}) \langle c^\dagger_i c^{\phantom\dagger}_j  \rangle$ with $\boldsymbol{\delta}\!=\!i\!-\!j$ (schematically).
We note that $\widetilde{D} \geq 0$, since it is the sum rule for ${\rm Re}\,\sigma(\omega) \geq 0$.
We then use the triangle inequality and Cauchy-Schwarz $|\langle c_i^\dag c_j^{\phantom\dag}\rangle| \leq \sqrt{\langle n_i \rangle\langle  n_j \rangle} = n$
to obtain $D_s \leq \widetilde{D} \leq  n \sum_{\boldsymbol{\delta}} \delta_a^2 | t(\boldsymbol{\delta}) |/2$. 
This shows that for small hopping and/or low density, one
necessarily has a small ${D}_s$.

{\bf $T_c$ bound in 2D:} 
For a BKT transition in 2D, the $T_c$ and the stiffness $D_s$ are related by the
universal ratio~\cite{NelsonPhys.Rev.Lett.1977} $k_B T_c/D_s\left(T_c^{-}\right) = \pi/2$.
Together with eq.~(\ref{eq:ineq}) $D_s\left(T_c^{-}\right) \leq \widetilde{D}\left(T_c\right)$,
we then immediately obtain eq.~(\ref{eq:2DTcbound}).
In an anisotropic system $\widetilde{D}$ depends on $a = x,y$ through the 
${\partial^2/\partial k_a^2}$ in eq.~(\ref{eq:mass}).
We can use $\widetilde{D} = \max\left\{\widetilde{D}_x,\widetilde{D}_y\right\}$
to obtain a bound on $T_c$, however, we argue in Appendix~\ref{sec:aniso}, 
for a much stronger result $\widetilde{D} = \left[\widetilde{D}_x\widetilde{D}_y\right]^{1/2}$ in 2D.

We emphasize that eq.~(\ref{eq:2DTcbound}) with $\widetilde{D}(T_c)$ on the RHS 
is sufficient to derive the rigorous results below.
However, to obtain the intuitively more appealing result $k_B T_c \le \pi \widetilde{D}(0)/2$,
we need to assume that $D_s(T)$ is a decreasing function of $T$,
so that $D_s\left(T_c^{-}\right) \leq D_s\left(0\right) \leq \widetilde{D}\left(0\right)$.

{\bf 2D Parabolic Dispersion:} 
Consider a single band with $\epsilon({\bf k}) = \hbar^2 k^2/2m$
with density $n$, so that the Fermi energy $E_F = \pi\hbar^2  n/m$
and arbitrary interactions that lead to pairing and superconductivity.
Then $M^{-1}({\bf k}) = m^{-1}$ and 
$\Omega^{-1}\sum_{{\bf k},\sigma} n_\sigma({\bf k}; T) = n$ independent of $T$ and interactions,
so that $\widetilde{D} = \hbar^2 n/4m$.
Eq.~(\ref{eq:2DTcbound}) then leads to the simple result
\begin{equation}
k_BT_c \leq E_F/8
\label{eq:Ef-over-8}
\end{equation}
which must be obeyed independent of the strength of attraction or order-parameter symmetry, provided
the system exhibits a BKT transition.
In a weak-coupling superconductor $T_c$ will actually be {\it much smaller} than
$E_F/8$ but, as we discuss next, the bound can be saturated 
in systems with strong interactions, such as the 2D BCS-BEC crossover experiments in ultra-cold Fermi gases.

{\bf 2D BCS-BEC crossover:} 
In ultra-cold Fermi gas experiments the two-body s-wave interaction between atoms  is tuned using a Feshbach resonance. 
This has led to deep insights into the crossover~\cite{ketterle.zweirlein.2006,RanderiaAnnu.Rev.Condens.MatterPhys.2014}
from the weak coupling BCS limit with large Cooper pairs all the way to the BEC of tightly bound diatomic molecules.
Asymptotically exact results are available in both the BCS and BEC limits, however,
the crossover regime between the two extremes is very strongly interacting, with pair size comparable to
the inter-particle spacing, and is much less understood. It is precisely here that our exact upper bound~(\ref{eq:Ef-over-8}) is relevant.

The 2D crossover for s-wave pairing is parameterized by the dimensionless interaction~\cite{RanderiaPhys.Rev.Lett.1989} $\log(E_b/E_F)$, 
where $E_b$ is the binding energy of the two-body bound state in vacuum and $E_F$ the Fermi energy. 
In the weak-coupling BCS limit ($E_b\!\ll\!E_F$), the mean field $k_B T_c \sim \sqrt{E_F E_b}$~\cite{RanderiaPhys.Rev.Lett.1989}, 
with a pre-factor that has been computed including the Gorkov-Melik-Barkhudarov (GMB) correction~\cite{GMB1961, PetrovPhys.Rev.A2003}. 
Clearly $T_c$ is much smaller than our bound.

In the BEC limit ($E_b\!\gg\!E_F$)  the composite bosons have mass $2m$, density $n/2$, and an inter-boson scattering length $a_b$ 
where $E_b/E_F \sim 1/na_b^2$~\cite{PetrovPhys.Rev.A2003}. The 2D dilute Bose gas has $k_B T_c =  E_F/[2\log\log(2/na_b^2)]$~\cite{FisherPhys.Rev.B1988}, 
which is valid in the regime $\log\log \gg 1$. This too is smaller than our bound, though 
our exact result cautions against a naive extrapolation of the BEC limit 
result into the strong interaction regime.

The results of the 2D Fermi gas experiment of ref.~\cite{RiesPhys.Rev.Lett.2015} seems to violate eq.~(\ref{eq:Ef-over-8})
in the crossover regime.
We note, however, that our bound is obtained for a strictly 2D system in the thermodynamic limit,
while the experiment is on a quasi-2D system in a harmonic trap, from which it is difficult to accurately determine the BKT $T_c$.  
The finite size of the trap raises $T_c$; even the non-interacting 
Bose gas in a 2D harmonic trap has a non-zero $T_c$.

{\bf Magic angle twisted bilayer graphene:} 
Let us next turn to a multi-band system of great current interest.
The existence of very narrow bands in MA-TBG was predicted by
continuum electronic structure calculations~\cite{BistritzerProc.Natl.Acad.Sci.2011,LopesdosSantosPhys.Rev.Lett.2007}
that pointed out the crucial role of $\alpha = w/\hbar v_F^0 K\theta$, where
$\theta$ is the twist angle between the two layers, $w$ is the interlayer tunneling, 
$v_F^0$ the bare Fermi velocity, and $K$ the Dirac-node location in monolayer graphene.
It was predicted that $v_F$ in TBG can be tuned to zero~\cite{BistritzerProc.Natl.Acad.Sci.2011}, with a bandwidth less than 10 meV
by choosing certain magic angles $\theta$, the largest of which $\approx 1.1^\circ$ has now been achieved in 
experiments~\cite{CaoNature2018a,CaoNature2018}.
Recently, pressure-tuning of $w$ has also resulted in very narrow bands~\cite{YankowitzScience2019}.

Little is known at this time about the nature of the SC state or the pairing mechanism,
though the observed non-linear I-V characteristics~\cite{CaoNature2018a,CaoNature2018} are consistent with a BKT transition.
Proximity to a ``Mott'' insulator and narrow bandwidth suggest the importance of electron correlations,
while the extreme sensitivity of the dispersion to structure suggests that electron-phonon
interactions could also be important. We argue here that simply using the available
electronic structure information for MA-TBG, and without any prejudice about the interactions
responsible for SC, we can put strong constraints on its superconducting $T_c$. 

There are two bands for each of the two valleys, one above and the other below the charge neutrality point (CNP) .
Each band has a two-fold spin degeneracy, with bands for one valley related to those of the other by time-reversal.
We include these eight bands in the $\sum_{mm',\sigma}$ in eq.~(\ref{eq:ineq}), while the $\sum_{{\bf k}}$ is over the moir\'e Brillouin zone, 
a hexagon with side $2K\sin(\theta/2) \simeq K\theta$.
We use the tight-binding model of ref.~\cite{KoshinoPhys.Rev.X2018a},
a multi-parameter fit to the continuum dispersion~\cite{BistritzerProc.Natl.Acad.Sci.2011},
to calculate $M^{-1}_{m,m'}({\bf k})$ of eq.~(\ref{eq:mass}), which is block-diagonal in the valley index,
so that there are no cross-valley terms in eq.~(\ref{eq:ineq}).

To derive a general bound, where we make no simplifying assumptions, we start with
$\widetilde{D}\!\geq\!0$ and obtain
$\widetilde{D} \leq (\hbar^2/4\Omega) \sum_{{\bf k}mm'\sigma} \vert M^{-1}_{mm'}({\bf k})\vert
\vert \langle c^\dag_{{\bf k}m\sigma}c^{\phantom\dag}_{{\bf k}m'\sigma} \rangle\vert$
using the triangle inequality. We next use Cauchy-Schwarz to obtain
$\vert \langle c^\dag_{{\bf k}m\sigma}c^{\phantom\dag}_{{\bf k}m'\sigma} \rangle\vert^2 \leq n_{m\sigma}({\bf k})n_{m'\sigma}({\bf k}) \leq 1$,
since the momentum distribution $n_{m\sigma}({\bf k}) \leq 1$.
We thus find $\widetilde{D} \leq (\hbar^2/4\Omega) \sum_{{\bf k},m,m'\sigma}\vert M^{-1}_{mm'}({\bf k})\vert$
which leads to the bound $k_B T_c \leq 56$ K.

We can obtain a more stringent $T_c$ bound if use further physical inputs. The ``Mott'' gap in the correlated insulator is 
experimentally~\cite{CaoNature2018a,CaoNature2018} known be $\approx 0.3$ meV, and we expect 
a superconducting gap which is at most that value. Thus we may assume that, at half-filling away from CNP on the
hole doped side, say, the bands above the CNP are essentially empty and unaffected by pairing.

Before proceeding, we derive a general result valid for arbitrary interactions which shows that
inter-band terms do not contribute to eq.~(\ref{eq:ineq}) for completely filled or empty bands.
To prove this, we again use the Cauchy-Schwarz inequality
$\vert \langle c^\dag_{{\bf k}m\sigma}c^{\phantom\dag}_{{\bf k}m'\sigma} \rangle\vert^2 \leq n_{m\sigma}({\bf k})n_{m'\sigma}({\bf k})
= 0$ when either band $m$ or $m'$ is empty.
A similar argument works for the filled case after a particle-hole transformation; see Appendix~\ref{sec:offdiagonal}.
Thus $\langle c^\dag_{{\bf k}m\sigma}c^{\phantom\dag}_{{\bf k}m'\sigma} \rangle\!=\!0$ for $m\!\neq\!m'$,
whenever either of the two bands is completely filled or empty,
and only $m\!=\!m'$ terms survive in eq.~(\ref{eq:ineq}).

To bound $T_c$ for MA-TBG near half-filling on the {\it hole-doped side} of the CNP,
we take $n_m({\bf k})\!=\!0$ for the empty bands above the CNP, as explained above.
Keeping only band-diagonal terms and using the  triangle inequality
we obtain
$\widetilde{D} \leq (\hbar^2/4\Omega) \sum_{{\bf k},m,\sigma}\vert M^{-1}_{mm}({\bf k})\vert n_{m\sigma}({\bf k})$.
Using $n({\bf k}) \leq 1$ for the bands below CNP we 
obtain the bound $T_c \leq 14.4$ K near half-filling for hole doping
using the tight-binding model of ref.~\cite{KoshinoPhys.Rev.X2018a}.
A similar calculation leads to $T_c \leq 15.0$ K near half-filling for electron doping; see Appendix~\ref{sec:MATBG}.
We note that using $\left\vert M^{-1}\right\vert$ and general constraints on $n({\bf k})$
leads to rigorous results, but weakens the bounds.

\begin{figure}
\centering
\includegraphics[width=0.45\textwidth]{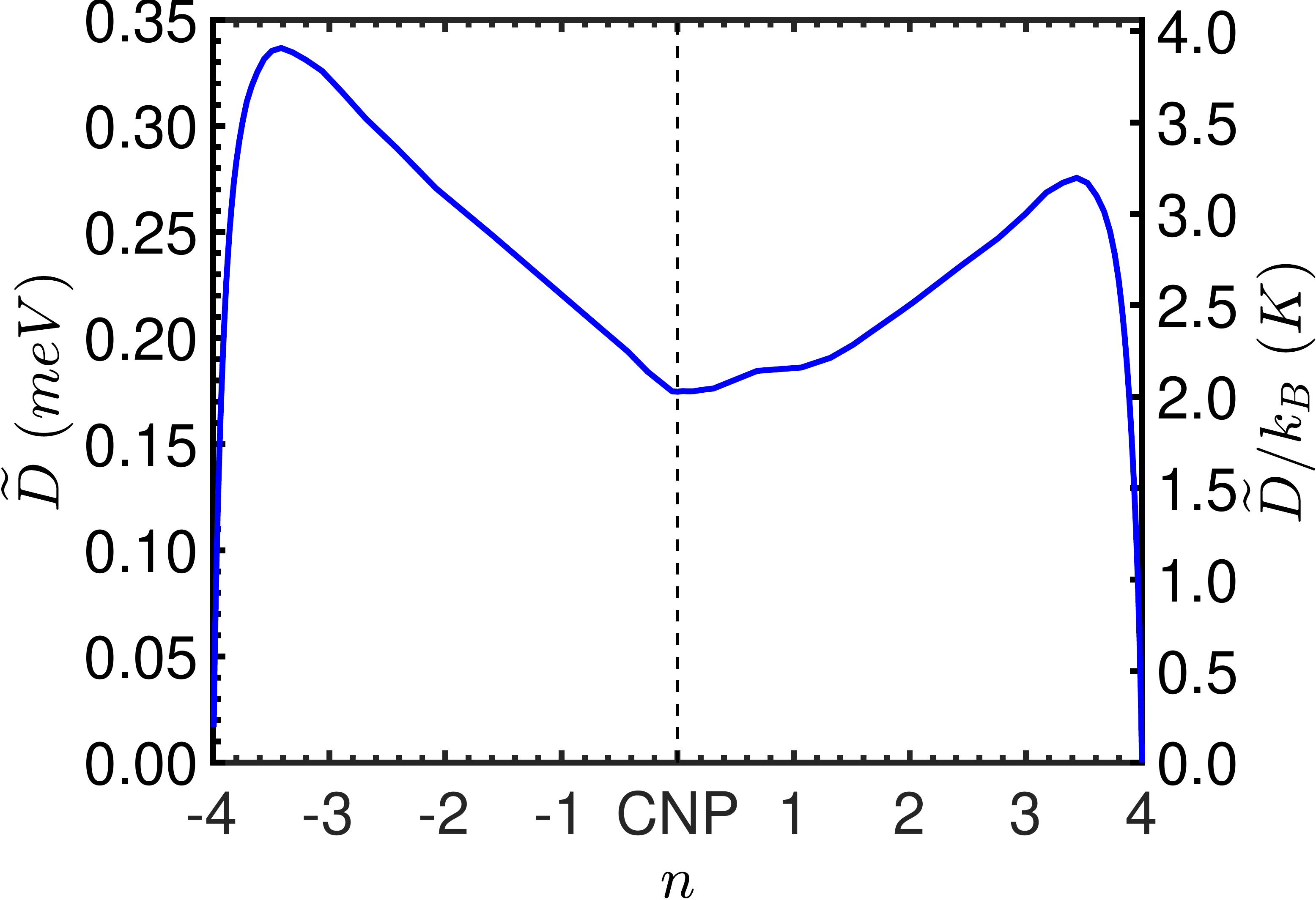}
\caption{  
$\widetilde{D}$ as a function of doping from the charge neutrality point (CNP) in magic angle-twisted bilayer graphene (MA-TBG), 
calculated using the band structure of ref.~\cite{KoshinoPhys.Rev.X2018a} at $T\!=\!0$. 
$\left(2\pi e^2/\hbar^2\right)\widetilde{D}$ is the integrated optical spectral weight and
$\pi\widetilde{D}/2$ is an upper bound on the SC $T_c$ in MA-TBG.
}
\label{fig:tbg}
\end{figure}

\begin{figure}
\centering
\includegraphics[width=0.4\textwidth]{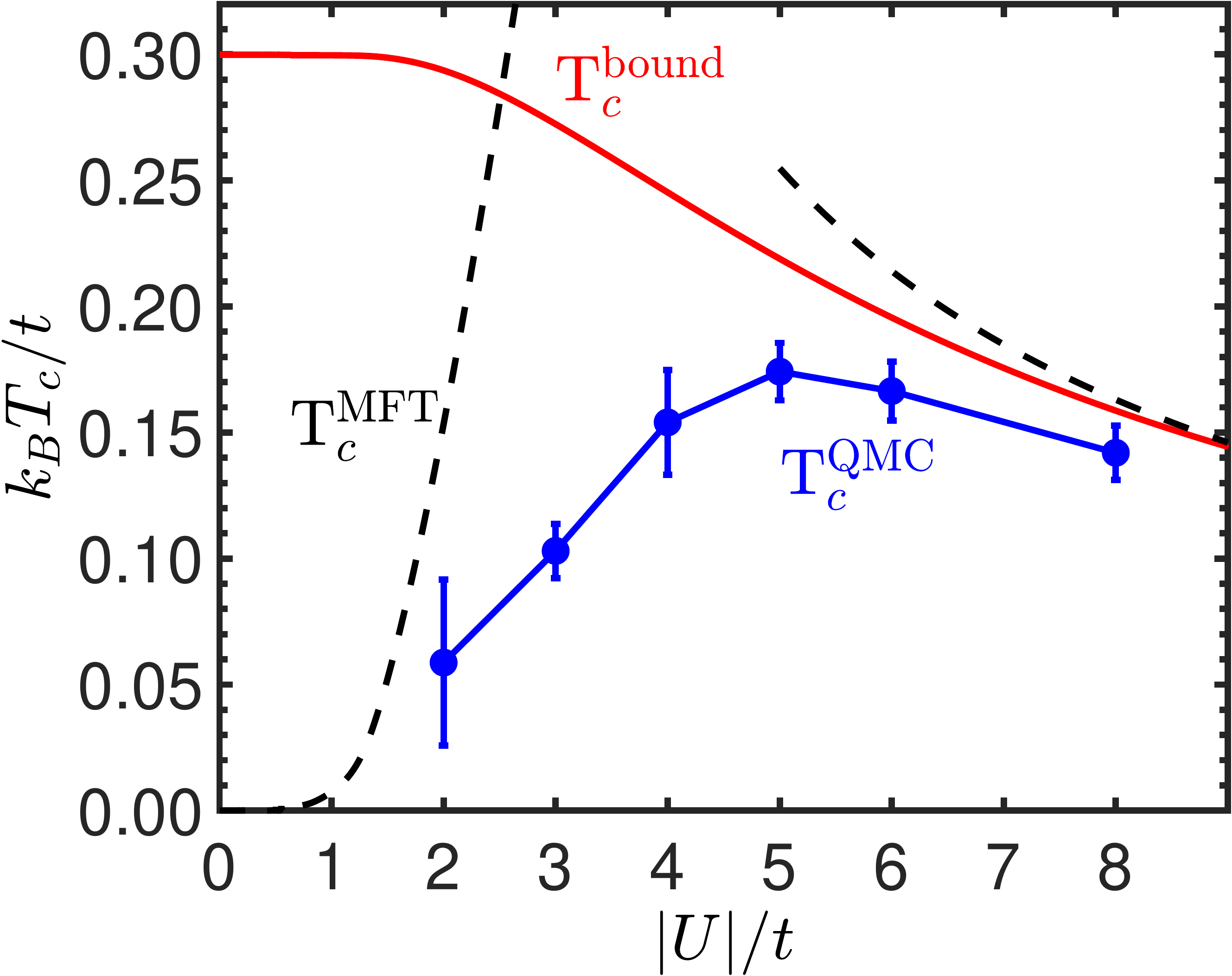}
\caption{$T_c$ for the 2D attractive Hubbard model at density $n\! =\!0.7$ with
QMC results from ref.~\cite{PaivaPhys.Rev.Lett.2010a}.
The BCS mean field $T_c^{\rm MFT}$ controls $T_c$ at weak coupling. 
Phase fluctuations, estimated using our upper bound $T_c^{\rm bound}$ (see text)
dominate at intermediate and strong  coupling, where we also 
show the $t^2/|U|$ asymptotics of our bound.}
\label{fig:negu}
\end{figure}

Finally, we make a physically motivated estimate of $\widetilde{D}$, which yields an improved, but approximate, result. 
We use the $T\!=\!0$ band theory result 
$\langle c^\dag_{{\bf k}m\sigma}c^{\phantom\dag}_{{\bf k}m'\sigma} \rangle = \delta_{m,m'}\Theta\left(\mu - \epsilon_m({\bf k})\right)$,
with the chemical potential $\mu$ determined by the density $\Omega^{-1} \sum_{{\bf k},m,\sigma} n_{m\sigma}({\bf k})$.
This, together with $M^{-1}_{mm}({\bf k})$ calculated from the tight binding model of ref.~\cite{KoshinoPhys.Rev.X2018a},
leads to the density-dependent estimate of $\widetilde{D}$ plotted in Fig.~\ref{fig:tbg}. 
We note that using $\partial^{2} / \partial k_{x}^{2}$ versus $ \partial^{2} / \partial k_{y}^{2}$ to calculate $M^{-1}$ 
affects our estimates by less than a percent.

The integrated optical spectral weight, given by $\left(2\pi e^2/\hbar^2\right)\widetilde{D}$,
vanishes at the band insulators when all bands are either filled or empty. Clearly our band-structure based
estimate does not know about the ``Mott" insulating states at half-filling away from CNP.
$(\pi/2)$ times the $\widetilde{D}$ plotted in Fig.~\ref{fig:tbg} is an estimated upper bound on the SC $T_c$. 
The system is not SC over most of the doping range, but our bound is the
maximum attainable $T_c$ if the system were to exhibit superconductivity.
We find the maximum $T_c$ to be about 6 K, while the experimental value is 3 K~\cite{YankowitzScience2019}.

We note that the $T_c$ bounds are sensitive to the precise electronic structure results we use as input
for calculating $M^{-1}$. As shown in Appendix~\ref{sec:MATBG}, using the tight binding results of  ref.~\cite{KangPhys.Rev.X2018} 
for MA-TBG, leads to a $T_c$ estimate about 2.5 times higher than the one presented above, based on the band structure of ref.~\cite{KoshinoPhys.Rev.X2018a}. We emphasize that these differences arise from the fact that the details of the 
non-interacting band structure of MA-TBG are not very well established. Irrespective of that, our results suggest that MA-TBG is a 
strongly correlated SC in a phase fluctuation dominated regime.

{\bf 2D attractive Hubbard model and optical lattices:} 
We next obtain important insights on the value of $T_c$ and its interaction-dependence
for the 2D attractive Hubbard model, where we can compare our bound with 
sign problem free Quantum Monte Carlo (QMC) simulations~\cite{PaivaPhys.Rev.Lett.2010a}.
This system has also been investigated in recent optical lattice 
experiments~\cite{MitraNaturePhys2018}.

Consider nearest-neighbor (NN) hopping on a square lattice with
$\mathcal{H} = -t \sum_{\langle i,j \rangle \sigma}c^\dagger_{i,\sigma} c^{\phantom{\dag}}_{j\sigma} + {\rm{h.c.}}- |U| \sum_{i} 
\left(n_{i \uparrow} -1/2\right)\left(n_{i\downarrow}-1/2\right)$. For $n\!\neq\!1$ the system
has an s-wave SC ground state, exhibiting a crossover from a weak coupling BCS state ($|U|/t \ll 1$) to a BEC of hard-core on-site bosons
($|U|/t \gg 1$). 
The QMC estimate~\cite{PaivaPhys.Rev.Lett.2010a} of $T_c$, obtained from the BKT jump in the
$D_s$, is a non-monotonic function of $|U|/t$ at a fixed density $n$; see Fig.~2.
The BCS mean field $T_c^{\rm MFT}$ correctly describes the weak coupling $T_c$, 
(For a more accurate estimate, one should take into account the GMB correction~\cite{GMB1961} 
which suppresses the numerical pre-factor, but does not alter the functional form of  $T_c^{\rm MFT}$.)
For $|U|/t > 2$, $T_c^{\rm MFT}$ is the scale at which pairs dissociate and lies well above $T_c$. In the
$|U|/t \gg 1$  limit we see $T_c \sim t^2/|U|$, the effective boson hopping.

Our bound permits us to understand $T_c(|U|/t)$ in the intermediate coupling
regime where there are no other reliable analytical estimates.  
To estimate $\widetilde{D}$ analytically, 
we need to make an approximation for $n({\bf k})$. If we choose a step-function
(as we did for the MA-TBG) we get $T_c\!\leq\!0.3 t$ for $n\!=\!0.7$, independent of $|U|/t$.

To obtain a better estimate, we note that, as $|U|/t$ increases, the pair-size shrinks and $n({\bf k})$ broadens. 
In the extreme $|U|/t$-limit of on-site bosons, $n({\bf k})$ is flat (${\bf k}$-independent),
leading to $\widetilde{D}\!\rightarrow\!0$, since ${\partial^2 \epsilon}/{\partial k_x^2}$ is a periodic function 
with zero mean whose ${\bf k}$-sum vanishes.
To model this broadening of $n({\bf k})$, we use the results of the $T\!=\!0$ BCS-Leggett crossover theory; see Appendix~\ref{sec:negu}.
This gives us the (approximate) bound plotted in Fig.~2, which has the correct $t^2/|U|$ asymptotic behavior
at large $|U|$. 

In general, we see that $T_c \leq \min\left\{T_c^{\rm MFT}, \pi\widetilde{D}/2k_B\right\}$.
For temperatures between the pairing scale $T_c^{\rm MFT}$ and $T_c$ at which phase coherence sets in,
the  ``normal state'' exhibits a pseudogap due to pre-formed pairs~\cite{RanderiaPhys.Rev.Lett.1992,TrivediPhys.Rev.Lett.1995}.

{\bf Three dimensional systems:} 
Experiments suggest that there may be an upper bound on $T_c$ in 3D systems; 
see, e.g., Fig.~6 of ref.~\cite{CaoNature2018}.
We have not succeeded in deriving a rigorous bound on the 3D $T_c$, unlike in 2D. There are 
two challenges that one faces in trying to derive a bound in 3D, one related to rigorous 
control on numerical pre-factors and the other to the functional form of the relation between $T_c$ and $D_s$.
Both are related to the fact that in 3D the superfluid stiffness does not have dimensions of energy, unlike in 2D.

Following Emery and Kivelson (EK)~\cite{EmeryNature1995}, we focus on the 3D phase ordering temperature
$k_B T_\theta = A D_s(0)\, \overline{a}$, which could provide a bound on $T_c$.
Here $A$ is a (dimensionless) constant and $\overline{a}$ is the length-scale up to which one has to
coarse-grain to derive an effective XY model. 
EK use $\overline{a}^2 = \pi \xi^2$, where $\xi$ is the coherence length,
and suggest, based on Monte Carlo results for classical XY models,  that $A \simeq 4.4$ gave a reasonable account of
experiments on underdoped cuprates and other materials.

However, the coefficient $A$ is {\it non-universal} and can vary from one system to another. Consider the 3D
problem of the BCS-BEC crossover in ultra-cold Fermi gases~\cite{RanderiaAnnu.Rev.Condens.MatterPhys.2014} 
with $\hbar^2k^2/2m$ dispersion and interaction, characterized by the s-wave scattering length $a_s$, tuned using a Feshbach resonance. 
At unitarity ($|a_s| = \infty$), the experimental $k_B T_c \simeq 0.17 E_F$~\cite{KuScience2012},
while QMC estimates~\cite{BurovskiPhys.Rev.Lett.2008,GoulkoPhys.Rev.A2010} range from $k_B T_c \simeq 0.15 E_F - 0.17 E_F$.
QMC shows the expected non-monotonic behavior of $k_B T_c/E_F$ as a function of $1/k_F a_s$, with a maximum $k_B T_c/E_F\simeq 0.22$ 
at a small positive $1/k_F a_s$.
The maximum value of $k_B T_c/E_F$ is {\it larger} than the non-interacting BEC result, consistent
with the rigorous result~\cite{SeiringerPhys.Rev.B2009} that repulsive interactions increase the $T_c$ of a dilute Bose gas in 3D.

We choose $\xi \simeq k_F^{-1}$ near unitarity~\cite{EngelbrechtPhys.Rev.B1997} 
and try to use $k_B T_\theta = A (\hbar^2 n/4m) (\sqrt{\pi}\xi)$ as a bound on $T_c$. Consistency with the observed
$k_B T_c/E_F\simeq 0.22$ then requires $A \simeq 7.4$, quite different from the $4.4$ quoted above. 
We do not know if there is a definite value of $A$  that would give a ``phase-ordering" upper bound on $T_c$ in 3D.

The following argument suggests that there may, in fact, be no {\it general} bound on $T_c$ that is linear in $D_s(0)$ in 3D.
From a practical point of view, one is interested in learning about the highest $T_c$ in a class of materials. But, if a general bound 
were to exist, it should be equally valid in situations where both $T_c$ and $ D_s(0)$ are driven to zero by tuning 
a (dimensionless) parameter $\delta \rightarrow 0^{+}$ toward a quantum critical point (QCP). From the action 
$S = {1\over 2}D_s\int_0^\beta d\tau \int d^d{\bf r}|\nabla\theta|^2+ \ldots$ describing the phase fluctuations of
the SC order parameter, we get the quantum Josephson scaling relation~\cite{FisherPhys.Rev.B1989} $D_s(0) \sim \delta^{(z+d-2)\nu}$.
One also obtains, as usual, $T_c \sim \delta^{z\nu}$, where $z$ and $\nu$ are the dynamical and correlation length
exponents in $d$ spatial dimensions.
Thus $T_c \sim [D_s(0)]^{z/(z+d-2)}$ near the QCP. In 2D, this gives a linear scaling between $T_c$ and $D_s(0)$.
However, in 3D we get  $T_c \sim D_s(0)^{z/(z+1)}$ which, sufficiently close to the QCP, will necessarily violate
an upper bound on $T_c$ that is conjectured to scale linearly with $D_s(0)$.
This is not just an academic issue, as experiments see precisely such a deviation from linear scaling
with $T_c \sim \sqrt{D_s(0)}$, consistent with $z=1$, both in highly underdoped~\cite{HetelNat.Phys.2007,BrounPhys.Rev.Lett.2007}  
and in highly overdoped~\cite{LembergerPhys.Rev.B2011,BozovicNature2016} cuprates.

{\bf Concluding remarks:} 
We have thus far ignored disorder. We note that $D_s$ of the pure system is necessarily larger than that 
in the disordered system. This can be seen by generalizing Leggett's bound~\cite{Leggett1970PRL} on the superfluid density 
(derived in the context of supersolids) to the case of disordered systems~\cite{ParamekantiPhys.Rev.B1998}.
Thus our upper bounds for translationally invariant systems continue to be valid in the presence of disorder, although they can be improved.

Although we have focused on narrow band and low density systems here, our
bounds have also important implications for systems close to insulating states, either correlation-driven or disorder-driven.
In either case, if there is a continuous superconductor to insulator transition, the superfluid stiffness will eventually become smaller than
the energy gap and control the SC $T_c$. 

As a design principle, it is interesting to ask if one can have multi-band systems where a narrow band has a large energy gap and 
large ``mean field" $T_c$ interacting with a broad band that makes a large contribution to the superfluid stiffness, thus getting the best of both worlds.

\bigskip

\textbf{Acknowledgments} We are grateful to P. T\"orm\"a, S. Peotta and L. Liang for
pointing out an error in an earlier version of our paper that led us to the correct multi-band result presented here.
We thank J. Kang and O. Vafek for providing the tight-binding parameters for ref.~\cite{KangPhys.Rev.X2018}.
 We acknowledge support from NSF DMR-1410364 and the Center for Emergent Materials, an NSF MRSEC, under
Award Number DMR-1420451.

\bibliography{Tcboundsv10}

\newpage

\clearpage

\appendix

\section{Linear response, $D_s$ and $\widetilde{D}$}\label{sec:kubo}

Let us consider the general Hamiltonian
\begin{equation}
	  {\cal H} = {\cal H}_K +  {\cal H}_{\rm int}
	   \label{ham-full}
\end{equation}
where 
${\cal H}_{\rm int}$ represents arbitrary interactions, including those that gives rise to superconductivity, and 
$\mathcal{H}_{\rm K}$ is the most general single particle Hamiltonian for a multi-band/multi-orbital lattice model 
\begin{equation}
	  {\cal H}_{\rm K} = \sum\limits_{ i \alpha j \beta \sigma} t_{\alpha\beta}( {\bf r}_{i\alpha} - {\bf r}_{j\beta} ) c^{ \dagger }_{ i \alpha } c^{ \phantom{\dagger} }_{ j \beta  } .\label{ham-general}
\end{equation}
Here $t_{\alpha\beta}( {\bf r}_{i\alpha} - {\bf r}_{j\beta} ) $ represents the hopping matrix element 
from orbital $\beta$ in unit cell $j$ to orbital $\alpha$ in unit cell $i$ with $i,j$ spanning all unit cells, including $i=j$.
{\it We omit the spin label $\sigma$ only to simplify notation but we are not ignoring spin}, as emphasized by the spin sum.
In the presence of an external vector potential ${\bf A}$, the hopping picks up the Peierls phase 
\begin{equation}
	 {\cal H}_{\rm K} \; \rightarrow \;{\cal H}_{\rm K} = \sum\limits_{ {\bf R} {\bf r}, \alpha \beta \sigma } t_{\alpha\beta} ( {\bf   r} ) e^{ -i e {\bf A} ( {\bf R} ) \cdot {\bf   r}/\hbar }  c^{ \dagger }_{ i \alpha  } c^{ \phantom{\dagger} }_{ j \beta  }
\end{equation}
where we use the notation ${\bf R} = ({\bf r}_{i\alpha} 
+ {\bf r}_{j\beta})/2 $ and ${\bf   r} = {\bf r}_{i\alpha} - {\bf r}_{j\beta}$ for simplicity.
Since we are eventually interested in the long wavelength limit ${\bf q} \rightarrow 0$,
we choose a very slowly varying vector potential and write $ \int\limits_{ {\bf r}_{ j\beta } }^{ {\bf r}_{ i\alpha} } {\bf A} \cdot d {\bf l} \simeq {\bf A}( {\bf R} ) \cdot {\bf   r} $.

Within linear response theory we can Taylor expand the exponential retaining
terms which are linear (paramagnetic) and quadratic (diamagnetic) in ${\bf A}$. 
We transform to Fourier space using $t_{\alpha\beta}( {\bf k } ) = \sum_{ {\bf r} } t_{\alpha\beta} ( {\bf   r} ) e^{- i {\bf k} \cdot  {\bf   r} }$
and $c_{ i \alpha  } = \Omega^{-1/2} \sum_{ {\bf k} } e^{ i {\bf k} \cdot {\bf r}_{i\alpha } }  d_{ {\bf k} \alpha  } $.
We can then write the current operator ${j}_x = \delta  \mathcal{H}_{\rm K}/\delta A_x$ as the sum of the
paramagnetic $(P)$ and diamagnetic $(D)$ current operators given by
\begin{eqnarray}
{j}^P_x({\bf q}) &=& \frac{e}{\hbar\Omega} \sum\limits_{ \alpha\beta,  {\bf k}  \sigma }  \frac{\partial t_{\alpha\beta}( {\bf k } )}{\partial k_x}  d^{ \dagger }_{ {\bf k}+{\bf q}/2, \alpha  } d^{ \phantom{\dagger} }_{ {\bf k}-{\bf q}/2, \beta  }\\
{j}^D_x({\bf q}) &=& \frac{e^2}{\hbar^2 \Omega}   \sum\limits_{ \alpha \beta, {\bf k} \sigma} \frac{ \partial^2 t_{\alpha\beta}( {\bf k } )}{\partial k_x^2}  d^{ \dagger }_{ {\bf k} \alpha  } d^{ \phantom{\dagger} }_{ {\bf k} \beta  } A_x ( {\bf q} ),
\end{eqnarray}
where we only show the $x$-component for simplicity. Note that the paramagnetic current operator, when transformed to the band basis, will in general have
interband matrix elements~\cite{PeottaNatComm2015,LiangPhysRevB2017}. The only property of ${j}^P_x({\bf q})$ that we will need to
use below, however, is that it is a Hermitian operator; see equation~(\ref{eq:chi-bound}).

The superfluid stiffness $D_s$ is defined as the static long-wavelength limit of the transverse response of the current density ${\bf j}$ to a vector potential ${\bf A}$
\begin{eqnarray}
    \langle j_x \rangle({\bf q},\omega) &=& \frac{-4e^2}{\hbar^2} D_s A_x({\bf q},\omega) \nonumber \\
    & &\quad {\rm with} \quad q_x=0,q_\perp\rightarrow 0, \omega=0
\end{eqnarray}
and $\perp$ represents the orthogonal directions to $x$.
Standard linear response theory leads to the Kubo formula
\begin{eqnarray}
    D_s =  \widetilde{D} - \frac{\hbar^2}{4e^2} \chi_{j_x j_x}^\perp( {\bf q }\rightarrow 0 ,\omega =0)\label{eq:Kubo}
\end{eqnarray}
where the first term is the diamagnetic term, which is of central interest in this work, and the second is the transverse
paramagnetic current-current correlation function. We will focus on the latter in 
Appendix \ref{sec:para}, where we
show that $\chi_{j_x j_x}^\perp \geq 0$ at all temperatures.

Here we focus on the first term that can be read off from the form of the diamagnetic current operator. We find it convenient to 
write it in the band basis as
\begin{eqnarray}
\widetilde{D} = \frac{\hbar^2}{4\Omega} \sum\limits_{ m m^\prime, {\bf k} \sigma} M^{-1}_{mm^\prime}({\bf k}) 
\left\langle c^{ \dagger }_{ {\bf k} m  } c^{ \phantom{\dagger} }_{ {\bf k} m^\prime  } \right\rangle
\label{eq:tildeD}
\end{eqnarray} 
with the inverse mass tensor given by
\begin{eqnarray}
M^{-1}_{mm^\prime}({\bf k}) = \sum\limits_{\alpha\beta}\ 
U^\dag _{m,\alpha}({\bf k}) \ \frac{ \partial^2 t_{\alpha\beta}({\bf k})}{\partial(\hbar k_x)^2}\ U_{\beta,m^\prime}({\bf k}).
\end{eqnarray} 
The unitary transformation $U$ that transforms from the orbital to the band basis is defined by 
 \begin{equation}
\sum_{\alpha\beta} \ U^\dag_{m,\alpha}({\bf k})\ t_{\alpha\beta}({\bf k})\ U_{\alpha,m'}({\bf k}) = \epsilon_m({\bf k})\ \delta_{m,m'}.
\label{eq:unitary}
\end{equation}
This allows us to write the final result in the band basis using
\begin{equation}
\ d_{{\bf k}\alpha} = \sum_m \ U_{\alpha,m}({\bf k}) c_{{\bf k} m }.
\end{equation}

We note several important points about the inverse mass tensor $M^{-1}_{mm^\prime}({\bf k})$.
(i) It depends only on the bare band structure, and is independent of temperature and interactions,
(ii) it has both diagonal and off-diagonal terms in the band indices. and 
(iii) it is {\it not} simply related to the curvature of
the bands $\partial^2\epsilon_m({\bf k})/\partial k_x^2$, in contrast to the single-band case in equation~(\ref{eq:tildeD-oneband}).

The standard reference on the formalism for calculating the superfluid stiffness in lattice systems is Scalapino, White and Zhang (SWZ)~\cite{ScalapinoPhys.Rev.B1993}.
Our normalization conventions differ from them and, more importantly, they focus on the special case of a single band model 
with nearest-neighbor (NN) hopping on a square (or cubic) lattice.
Thus it may be useful for us to provide a ``dictionary'' relating our results to theirs.

In the single-band case our expression for $\widetilde{D}$ reduces to
\begin{equation}
\widetilde{D} = \frac{1}{4\Omega} \sum\limits_{{\bf k} \sigma} \frac{ \partial^2 \epsilon({\bf k})}{\partial k_x^2} n({\bf k})
\label{eq:tildeD-oneband}
\end{equation}
where the momentum distribution
\begin{equation}
n({\bf k}) = \left\langle c^{ \dagger }_{{\bf k}}  c^{ \phantom{\dagger} }_{ {\bf k}} \right\rangle.
\end{equation}
This result is valid for arbitrary one-band dispersion. For the special case of 
nearest-neighbor (NN) hopping on a square (or cubic) lattice, it is easy to see that 
the right hand side of equation~(\ref{eq:tildeD-oneband}) is proportional to the kinetic energy in the $x$-direction,
$\langle - K_x \rangle$ in the notation of SWZ.  Our result thus reduces to 
\begin{equation}
\widetilde{D} \rightarrow \langle - K_x \rangle/4.
\end{equation}
Finally, we note that our superfluid stiffness $D_s$ is related to that of SWZ by 
\begin{eqnarray}
D_s =(\hbar^2/4\pi e^2)\ D_s^{\rm SWZ}
\end{eqnarray}

\section{Relation between $\widetilde{D}$ and optical spectral weight}\label{sec:optical}

To see that $\widetilde{D}$ is proportional to the optical sum rule spectral weight, we identify the dynamical conductivity $\sigma(\omega)$ as the current response to an electric field ${\bf E}=-\partial_t {\bf A}$
\begin{eqnarray}
    i\omega~\sigma(\omega) = \left[  \chi_{j_x j_x}({\bf q}=0,\omega) -\frac{4e^2}{\hbar^2} \widetilde{D}  \right]
\end{eqnarray}
Using the Kramers-Kr\"onig relation
\begin{eqnarray}
    \omega~{\rm Im}~\sigma(\omega) = -\frac{2}{\pi} 
    {\rm P}\int\limits_0^\infty d\omega'~{\rm Re}~\sigma(\omega') \frac{\omega^2}{\omega'^2 - \omega^2}
\end{eqnarray}
and ${\rm Re} \chi_{j_x j_x}(\omega \rightarrow \infty) \rightarrow 0$, we obtain the sum rule for the optical conductivity as
\begin{eqnarray}
    \int\limits_0^\infty d\omega~{\rm Re}~\sigma(\omega) = \frac{2 \pi e^2}{\hbar^2} \widetilde{D}
\end{eqnarray}

\section{Derivation of Bound $D_s \leq \widetilde{D}$}\label{sec:para}

We show that $\chi_{j_x j_x}( {\bf q } ,\omega=0 )\ge 0$  at any temperature. This follows directly from its Lehmann representation 
\begin{eqnarray}
\dfrac{1}{Z} \sum\limits_{ij} \left[ \frac{e^{-\beta E_i} - e^{-\beta E_j}}{E_j - E_i } \right]   | \langle i |  j^{P}_{x}({\bf q}) | j \rangle |^2  \; \geq 0
\label{eq:chi-bound}
\end{eqnarray}
where $|i\rangle$ and $|j \rangle$ are exact eigenstates of the full Hamiltonian $\mathcal{H}$ in equation~(\ref{ham-full}) with eigenvalues $E_i,E_j$
and $Z={\rm Tr}[e^{-\beta \mathcal{H}}]$. 
The last inequality follows from  $(e^{-x} - e^{-y})/(y-x) \ge 0$. 
At zero temperature, this expression reduces to 
\begin{eqnarray}
\chi_{j_x j_x}( {\bf q } , \omega = 0 ) = 2 \sum_{i} \dfrac{ |\langle i | j^{P}_{x}({\bf q}) | 0 \rangle |^2}{E_i-E_0} \geq  0
\end{eqnarray} 
where $|0\rangle$ is the ground state. From equation~(\ref{eq:Kubo}), we thus conclude that 
\begin{eqnarray}
    D_s \le \widetilde{D}
\end{eqnarray}

\section{Real space bound on $\widetilde{D}$}\label{sec:realspace}

Except in the case of a single parabolic band, $\widetilde{D}$ depends in general on
both the $T$ and the interactions, since the thermal average in
$\left\langle c^{ \dagger }_{ {\bf k} m  } c^{ \phantom{\dagger} }_{ {\bf k} m^\prime  } \right\rangle$
is calculated using the full ${\cal H}$.
It is thus illuminating to derive an upper bound for $\widetilde{D}$ which shows that 
$\widetilde{D}$ must become small when the densities are low or if all the hopping parameters are small.
Such a bound for the single-band case with arbitrary dispersion was sketched in the paper. Here we
turn to the multi band case.

It is convenient to start with the real space representation
\begin{equation}
\widetilde{D} = \frac{ 1 }{ 4\Omega } \sum \limits_{ {\bf R } {\bf r}, \alpha \beta \sigma } r_x^2 \;  t_{\alpha\beta}({\bf r}) \left\langle  c^{ \dagger }_{ i \alpha} c^{ \phantom{\dagger} }_{ j \beta } \right\rangle.
\end{equation}
Here both forward and backward hopping are accounted for in $\sum_{ {\bf r} }$ with $t_{\alpha\beta}(|{\bf r}|) = t^*_{\beta\alpha}(|{\bf r}|)$.
Since $\widetilde{D} \geq 0$ we can use the triangle inequality. Further using the Cauchy-Schwarz inequality we get
\begin{eqnarray}
\widetilde{D} &\leq & \frac{ 1 }{ 4\Omega } \sum \limits_{ {\bf R } {\bf r}, \alpha \beta \sigma } r_x^2 \; \left|  t_{\alpha\beta}({\bf r}) \left\langle c^{ \dagger }_{ i \alpha } c^{ \phantom{\dagger} }_{ j \beta }  \right\rangle \right| \nonumber \\
&\leq & \frac{1}{4\Omega} \sum \limits_{ {\bf R } {\bf r}, \alpha \beta \sigma } r_x^2 \left| t_{\alpha\beta}({\bf r}) \right| \sqrt{ n^{\phantom{}}_{ i \alpha } n^{\phantom{}}_{ j \beta }    }
\end{eqnarray}
where $n^{\phantom{}}_{ i \alpha } = \langle c^{ \dagger }_{ i \alpha } c^{ \phantom{\dagger} }_{ i \alpha} \rangle $.

Here and below we define an inner product 
for operators $A,B$ in terms of the thermal expectation value
$\langle A^\dag B\rangle$, which allows us to use the Cauchy-Schwarz inequality $|\langle A^\dag B\rangle|^2 \leq \langle A^\dag A\rangle \langle B^\dag B\rangle$.

\section{Interband contributions to $\widetilde{D}$}\label{sec:offdiagonal}

We discuss here the conditions under which we can ignore the inter-band contributions to $\widetilde{D}$ given by
\begin{eqnarray}
    \widetilde{D} = \frac{\hbar^2}{4\Omega} \sum\limits_{ m m^\prime, {\bf k} \sigma} M^{-1}_{mm^\prime}({\bf k}) 
    \left\langle c^{ \dagger }_{ {\bf k} m  } c^{ \phantom{\dagger} }_{ {\bf k} m^\prime  } \right\rangle
    \label{eq:Dtildenotdiag}
\end{eqnarray}
This requires us to understand when $\left\langle c^{ \dagger }_{ {\bf k} m  } c^{ \phantom{\dagger} }_{ {\bf k} m^\prime  } \right\rangle =0$
for  $ m \neq m^\prime$.
We show here that this is the case, {\it independent of interactions}, when (a) either one of the two bands in empty, and (b) when either one of the two bands is fully filled.

We use the Cauchy-Schwarz inequality (see end of Appendix \ref{sec:realspace}) to obtain 
\begin{equation}
\left| \left\langle c^{\dagger}_{ {\bf k} m} c^{\phantom{\dagger}}_{ {\bf k} m^\prime} \right\rangle \right| \leq \sqrt{ n_{ m^{\phantom{}} } ( {\bf k} ) \; n_{ m^\prime } ( {\bf k} )   }
\end{equation}
where $n_{m}^{\phantom{\dagger}} ( {\bf k} ) = \left\langle c^\dagger_{ m {\bf k } } c^{\phantom{\dagger}}_{ m {\bf k } } \right\rangle$ is the momentum distribution function, and equality holds for $m=m^\prime$. For $m \neq m^\prime$, if either band is completely empty, $n_{m}^{\phantom{\dagger}} ( {\bf k} )=0$ for all ${\bf k}$ and the inter-band contribution to 
$\widetilde{D}$ in equation~(\ref{eq:Dtildenotdiag}) vanishes.

A similar argument for completely filled bands follows from a particle-hole transformation $c^{\phantom{\dag}}_{ m {\bf k} } \rightarrow h^{\dag}_{ m {\bf k} }$. Since $\left\langle c^{\dagger}_{ {\bf k} m} c^{\phantom{\dagger}}_{ {\bf k} m^\prime} \right\rangle = - \left\langle h^{\dagger}_{ {\bf k} m} h^{\phantom{\dagger}}_{ {\bf k} m^\prime} \right\rangle$,
\begin{eqnarray}
\left| \left\langle c^{\dagger}_{ {\bf k} m} c^{\phantom{\dagger}}_{ {\bf k} m^\prime} \right\rangle \right| &=& \left| \left\langle h^{\dagger}_{ {\bf k} m} h^{\phantom{\dagger}}_{ {\bf k} m^\prime} \right\rangle \right| \nonumber \\ 
&\leq & \sqrt{ n^h_{ m^{\phantom{\prime}} } ( {\bf k} ) \; n^h_{ m^\prime } ( {\bf k} )   } \nonumber \\
&=& \sqrt{ \left( 1- n_{ m^{\phantom{}} } ( {\bf k} ) \right) \;\left(1- n_{ m^\prime } ( {\bf k} )  \right) }.
\end{eqnarray}
Thus we conclude that for filled and empty bands, the inter-band terms do not contribute to the sum in equation~(\ref{eq:Dtildenotdiag}),
even in the presence of arbitrary interactions.

Finally, we note the simple fact that within band theory there are no inter-band contributions to $\widetilde{D}$.
In the absence of interactions (denoted by subscript $0$) we obtain
\begin{equation}
\left\langle c^{ \dagger }_{ {\bf k} m  } c^{ \phantom{\dagger} }_{ {\bf k} m^\prime  } \right\rangle_0 = f\left(\epsilon_m( {\bf k})\right)\delta_{m,m'}
\label{eq:non-int-cdag-c}
\end{equation}
where $f$ is the Fermi function.

\section{Magic Angle Twisted Bilayer Graphene (MA-TBG)}\label{sec:MATBG}

\begin{figure}
\centering
\includegraphics[width=8cm]{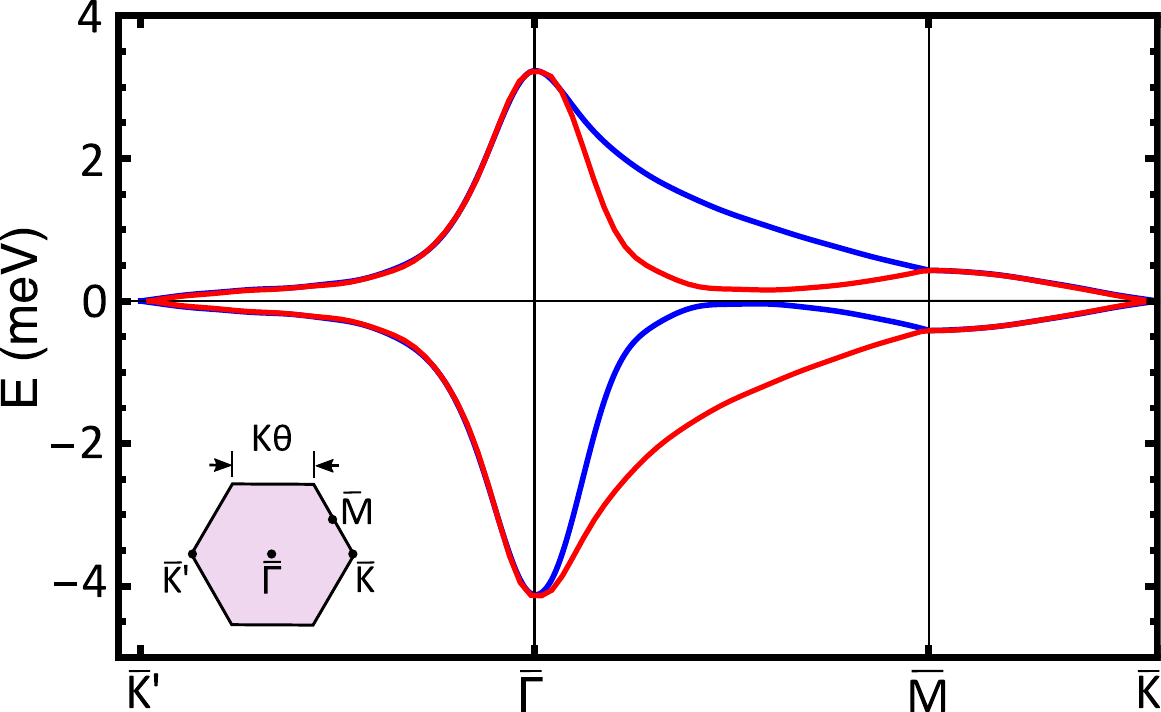}
\caption{Energy dispersion for MA-TBG along high-symmetry lines in the moire Brillouin zone (BZ) for the continuum model dispersion~\cite{BistritzerProc.Natl.Acad.Sci.2011} that is accurately 
described by the tight-binding model of Koshino \emph{et. al.}~\cite{KoshinoPhys.Rev.X2018a}.
The bands shown in red and blue correspond to the two valleys of the original BZ and are related by time reversal.
}
\label{FuPo}
\end{figure}

Magic angles in twisted bilayer graphene were first predicted by the continuum model~\cite{BistritzerProc.Natl.Acad.Sci.2011}.
Following up on the experimental discovery of correlation-induced insulators and superconductivity in MA-TBG,
there has been considerable progress in understanding its electronic structure~\cite{KoshinoPhys.Rev.X2018a,KangPhys.Rev.X2018,PoPhys.Rev.X2018}.
We first focus on the bounds that we obtain from the tight binding model of Koshino \emph{et.~al.}~\cite{KoshinoPhys.Rev.X2018a},
and then at the end of the Appendix compare these with the results we obtain from the tight binding model of 
Kang and Vafek~\cite{KangPhys.Rev.X2018}.

The continuum model dispersion~\cite{BistritzerProc.Natl.Acad.Sci.2011} is accurately reproduced by the multi-parameter tight binding fit 
of Koshino \emph{et.~al.}~\cite{KoshinoPhys.Rev.X2018a} (see Fig.~\ref{FuPo}) which takes into account hopping over 
distances up to $ 9 |{\bf L_M}|$ where ${\bf L_M}$ is the {\it moire} lattice vector. 
We use the hopping integrals presented in the Supplementary Information file \texttt{eff\_hopping\_ver2.dat} 
of ref.~\cite{KoshinoPhys.Rev.X2018a} to construct the non-interacting Hamiltonian $\mathcal{H}_K$ of equation~(\ref{ham-general}).
We then identify the unitary matrix $U({\bf k})$ that diagonalizes $t_{\alpha\beta}({\bf k})$ (see equation~(\ref{eq:unitary}))
and use it together with $ t_{\alpha\beta}({\bf k})$ to compute the inverse mass tensor 
\begin{eqnarray}
M^{-1}_{mm^\prime,a}({\bf k}) = \sum\limits_{\alpha\beta}\ 
U^\dag _{m,\alpha}({\bf k}) \ \frac{ \partial^2 t_{\alpha\beta}({\bf k})}{\partial(\hbar k_a)^2}\ U_{\beta,m^\prime}({\bf k}).
\end{eqnarray} 
Note that we have made explicit here the direction $a = x,y$ as an additional subscript on $M^{-1}$.

\begin{figure*}
\subfigure[\label{fig:Ekcomparison}]{
\includegraphics[height=130pt]{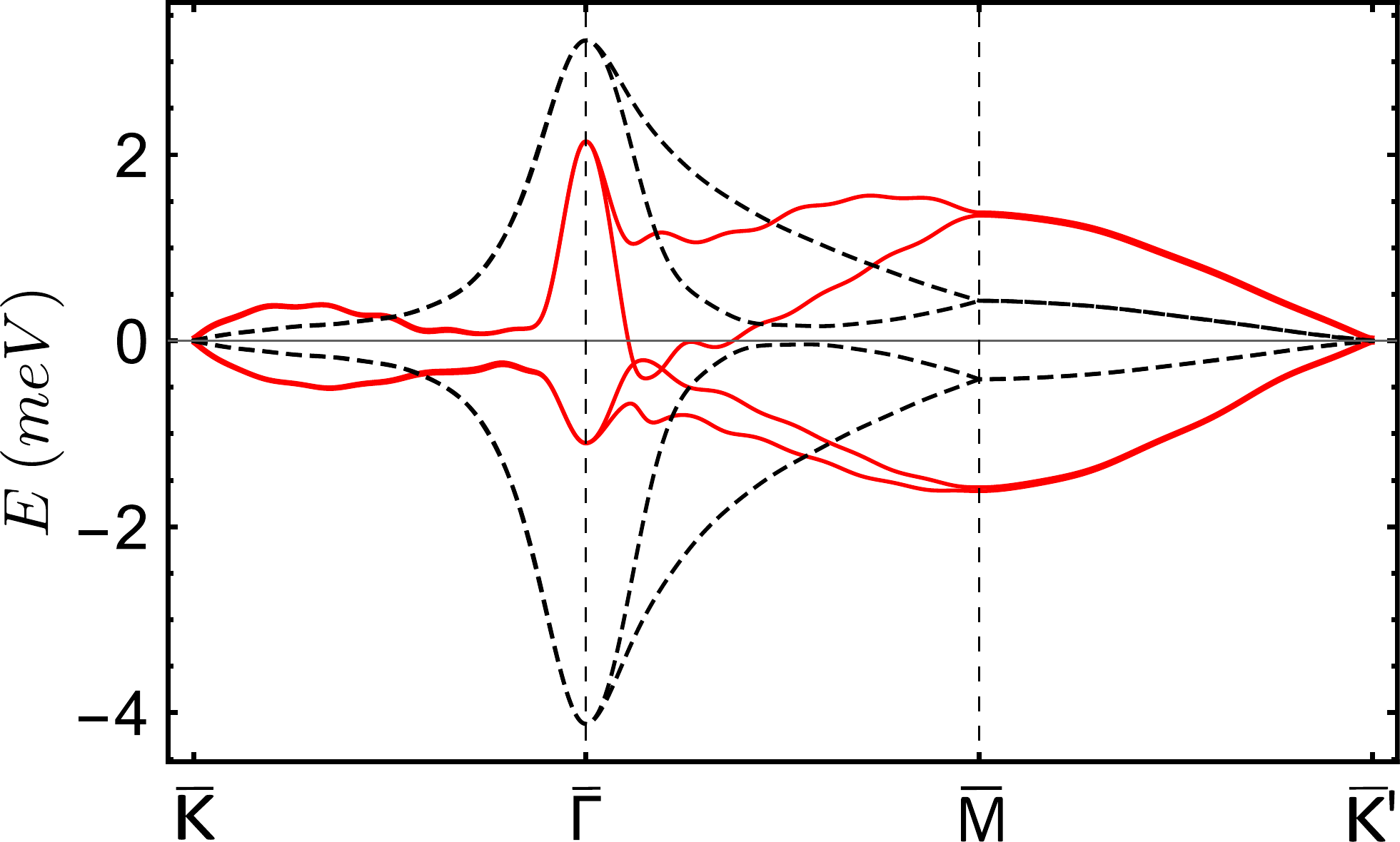}
}
\subfigure[\label{fig:Boundcomparison}]{
\includegraphics[height=130pt]{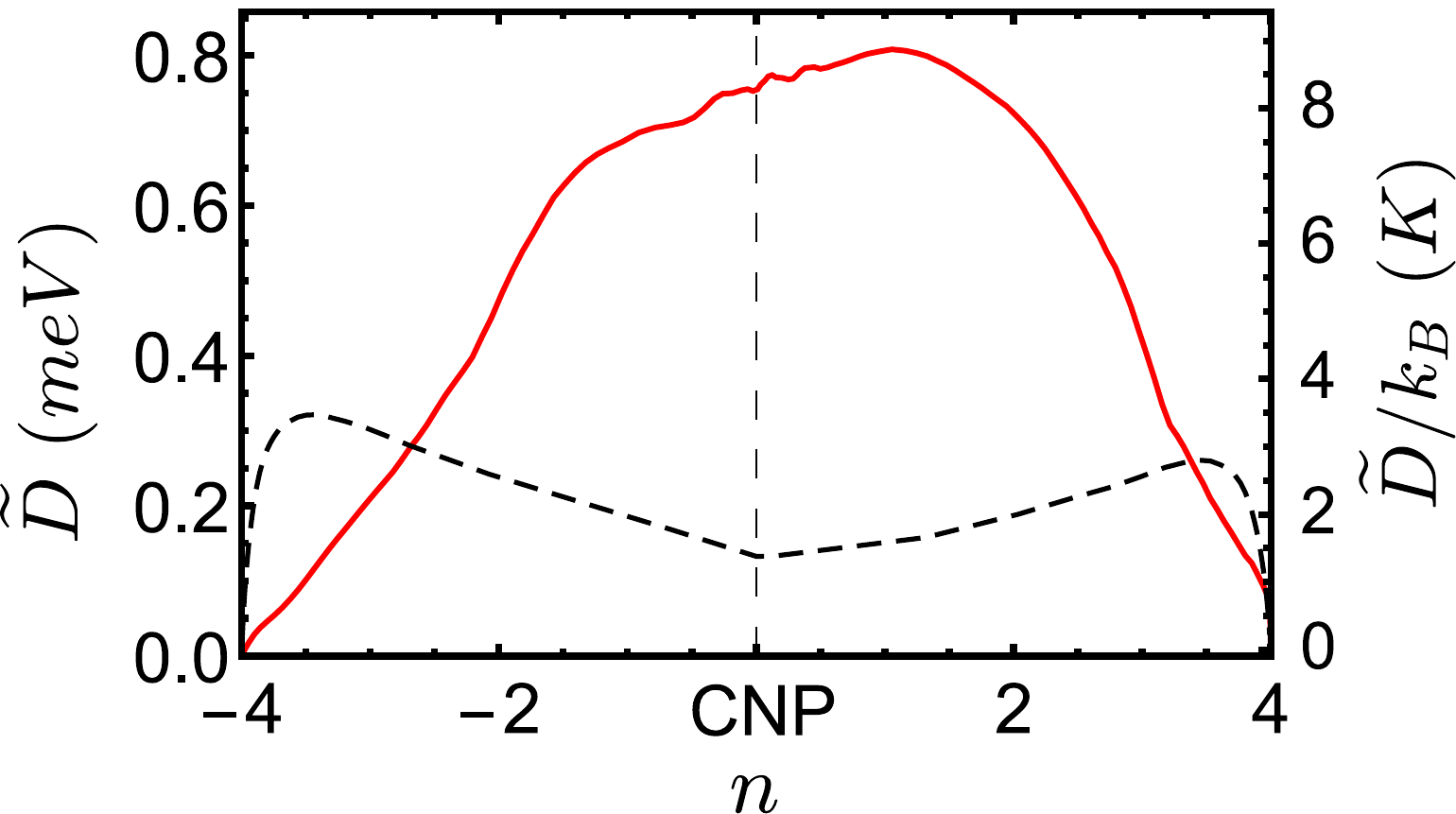}
}
\caption{\label{fig:KangVafekVsKoshinoFu}
Comparison of (a) the band structure and (b) the integrated spectral weight $\widetilde{D}$ for the models in ref.~\cite{KoshinoPhys.Rev.X2018a} (in black) and ref.~\cite{KangPhys.Rev.X2018} (in red).
}
\end{figure*}

The inverse mass tensor, obtained from the band structure information as described above, is used to compute  
$\widetilde{D}_x$ and $\widetilde{D}_y$ and bound $T_c$ as described in the paper.
The additional input needed to determine $\widetilde{D}$ using equation~(\ref{eq:tildeD}) is 
$\left\langle c^{ \dagger }_{ {\bf k} m  } c^{ \phantom{\dagger} }_{ {\bf k} m^\prime  } \right\rangle$, and 
we took two different approaches to compute this.

In the first approach, we looked at SC near half-filling on the hole-doped side of the CNP,
and argued that the chemical potential was sufficiently far from the CNP that we can take the band above
the CNP to be empty. Then using the result of Appendix \ref{sec:offdiagonal} we can ignore all inter-band terms with 
$m \neq m'$. For the occupied band we only used the general constraint that $n({\bf k}) \le 1$.
Using the triangle inequality, we then obtain
\begin{eqnarray}
    \widetilde{D}_a \le \frac{\hbar^2}{4\Omega} \sum\limits_{{\bf k}m,\sigma} \left| M^{-1}_{mm,a}({\bf k})  \right|.
\end{eqnarray}
where the empty bands above the CNP are excluded from the sum.

A similar reasoning also works for SC in the vicinity of half-filling on the electron-doped side of the CNP,
where we need to use the fact that the bands below CNP are filled to eliminate inter-band terms following 
Appendix \ref{sec:offdiagonal}.
We use a particle hole transformation
$c^{ \phantom{\dagger} }_{m{\bf k}} \rightarrow h^\dagger_{m{\bf k}}$,
under which $t_{\alpha\beta}({\bf k}) \rightarrow - t_{\alpha\beta}({\bf k})$
and thus $M^{-1}  \rightarrow - M^{-1}$.
We write $\widetilde{D}$ in terms of the hole momentum distribution functions
 $n^h_{m}({\bf k}) = \langle h^\dagger_{m{\bf k}} h^{ \phantom{\dagger} }_{m{\bf k}} \rangle$ to get
\begin{equation}
\widetilde{D}_a = {\hbar^2 \over 4\Omega} \sum\limits_{ m , {\bf k} \sigma} M^{-1}_{mm,a}({\bf k}) \left(n^h_m ({\bf k}) - 1 \right).
\end{equation}
We then show that the second term on the right hand side vanishes as follows:
\begin{eqnarray}
	\sum_{m,{\bf k}} M^{-1}_{mm,a}({\bf k}) &=& \sum_{{\bf k},\alpha\beta} {{\partial^2 t_{\alpha\beta}({\bf k})}\over{\partial(\hbar k_a)^2}} 
	\sum_m U^\dag _{m,\alpha}({\bf k}) U_{\beta,m}({\bf k}) \nonumber \\
	&=& \sum_{{\bf k},\alpha} {\partial^2 t_{\alpha\alpha}({\bf k}) \over\partial(\hbar k_a)^2} = 0.
\end{eqnarray}
We have first used $\sum_m U_{\beta,m}({\bf k}) U^\dag _{m,\alpha}({\bf k}) = \delta_{\beta,\alpha}$, which follows from the unitarity of $U$, 
and then the fact that ${\partial^2 t_{\alpha\alpha}({\bf k})/\partial k_a^2}$ is a periodic function with zero mean, whose
$\sum_{\bf k}$ vanishes. 
Using the triangle inequality and the general constraint $n^h({\bf k}) \le 1$, we obtain an expression for electron doping
which is similar to the hole-doped case:
\begin{equation}
    \widetilde{D}_a \le \frac{\hbar^2}{4\Omega} \sum\limits_{{\bf k}m,\sigma} \left| M^{-1}_{mm,a}({\bf k})  \right|
\end{equation}
where now the filled bands below the CNP are excluded from the sum.
These bounds, though rigorous, are weak because they involve $|M^{-1}|$ and only very general constraints on $n({\bf k})$.

The second (approximate) approach was to simply use a $T=0$ (non-interacting) band-theory estimate. 
We thus use equation~(\ref{eq:non-int-cdag-c}) to obtain
\begin{eqnarray}
    \widetilde{D}_a \simeq \frac{\hbar^2}{4\Omega} \sum\limits_{{\bf k}m,\sigma}  \ M^{-1}_{mm,a}({\bf k}) \Theta(\mu-\epsilon_{m}({\bf k}))
\end{eqnarray}
with the chemical potential $\mu$ determined by the density. 
We found that $\widetilde{D}_x$ and $\widetilde{D}_y$ calculated from the tight binding model of 
ref.~\cite{KoshinoPhys.Rev.X2018a} differ by less than a percent.
The resulting density-dependent $\widetilde{D}$ is shown in Fig.~1 of the main paper.

We note that there are many different tight binding models for describing the narrow bands in MA-TBG and our $T_c$ bounds depend
on this input. We have focused above on the results based on ref.~\cite{KoshinoPhys.Rev.X2018a} with an electronic structure that has separate
charge conservation at the $K$ and $K'$ valleys.  A rather different model without valley-charge conservation was 
derived~\cite{KangPhys.Rev.X2018} using only time-reversal and point group symmetry. We compare in Fig.~\ref{fig:Ekcomparison}
the band structures of ref.~\cite{KoshinoPhys.Rev.X2018a} in black and that of ref.~\cite{KangPhys.Rev.X2018} in red.
The corresponding integrated spectral weights $\widetilde{D}$ are shown in Fig.~\ref{fig:Boundcomparison} using the same
color convention. The maximum $T_c$ based on the band structure of ref.~\cite{KangPhys.Rev.X2018} is 15 K, which is 
2.5 times larger than that estimated from ref.~\cite{KoshinoPhys.Rev.X2018a}.

\section{Attractive Hubbard Model}\label{sec:negu}

It is interesting to ask how our bound on SC $T_c$ in 2D depends on interactions. We use the attractive Hubbard model on a square lattice as a 
concrete example to understand these trends, and to compare our bound with estimates of $T_c$ from sign-problem free quantum Monte Carlo
simulations.

Our bound is
$k_B T_c \le \pi/(8\Omega) \sum_{{\bf k},\sigma} \left( \partial_{k_x}^2 \epsilon({\bf k}) \right) n_{\sigma}({\bf k})$.
This result can be written in terms of the kinetic energy $\langle - K_x \rangle$ as discussed at the end of Appendix \ref{sec:kubo}.
The interaction-dependence is contained in the momentum distribution function $n_{\sigma}({\bf k})$ which, 
as we argued in the paper, must become increasingly broader and flatter as $|U|/t$ increases.
In the weak coupling BCS limit (small $|U|/t$) $n_{\sigma}({\bf k})$ is almost like the Fermi function at $T=0$, very slightly broadened by the superconductivity.
On the other hand in the extreme BEC limit (large $|U|/t$) of nearly on-site bosons, the $n_{\sigma}({\bf k})$ of the constituent fermions is essentially flat.

We model this $|U|/t$ trend in the momentum distributionn
using the BCS-Leggett crossover theory expression
\begin{eqnarray}
    n_{\sigma}({\bf k}) = \frac{1}{2} \left( 1- \frac{\epsilon({\bf k}) - \mu}{E({\bf k})} \right)
\end{eqnarray}
where $E({\bf k}) = \sqrt{(\epsilon({\bf k}) - \mu)^2 +\Delta^2}$ is the Bogoliubov quasiparticle energy.
The chemical potential $\mu$ and the pair potential $\Delta$ are  determined self-consistently for a given density $n$ and attraction $|U|$ by solving the 
$T\!=\!0$ gap and number equations
\begin{eqnarray}
    \frac{1}{|U|}=\frac{1}{\Omega} \sum_{{\bf k},\sigma} \frac{1}{2E({\bf k})} \\
    n=\frac{1}{\Omega} \sum_{{\bf k},\sigma} n_{\sigma}({\bf k})
\end{eqnarray}

We see from Fig.~2 that the $T_c$ obtained from QMC data~\cite{PaivaPhys.Rev.Lett.2010a} is always lower than $T_c^{\rm bound}$. Fig.~2 also shows that the bound is most useful in the intermediate to strong coupling regime, and less useful in the weak coupling regime where $T_c$ is, in fact, well
described by $T_c^{\rm MFT}$, the pair breaking energy scale.

\section{$T_c$ Bounds in spatially anisotropic systems}\label{sec:aniso}
\label{sec:anisotropy}

We collect here some results on the role of spatial anisotropy focusing mainly on 2D. We note that various quantities that 
we have considered are different in different directions labeled by $a = x,y$.
We have shown that 
\begin{equation} 
D_{s,a}(T) \leq \widetilde{D}_a(T).
\label{eq:aniso-bound}
\end{equation}
The most conservative bound on $T_c$ in 2D is then
\begin{equation}
k_B T_c \le {\pi \over 2}\ \max\left\{ \widetilde{D}_x,\widetilde{D}_y \right\}.
\end{equation}

Clearly this bound is not optimal because we expect $T_c$ to go to zero if either $D_{s,x}$ or $D_{s,y}$ goes to zero.
Using BKT theory we can show that
\begin{equation}
k_B T_c = {\pi \over 2}\ \left( D_{s,x}(T_c^-) D_{s,y} (T_c^-) \right)^{1/2} 
\label{eq:aniso-bkt}
\end{equation}
which leads to the improved bound
\begin{equation}
k_B T_c \leq {\pi \over 2}\ \left( \widetilde{D}_x \widetilde{D}_y \right)^{1/2} 
\end{equation}

To derive equation~(\ref{eq:aniso-bkt}) we start with the Free energy for phase fluctuations
\begin{equation}
{\cal F}= {1 \over 2}\ \int dx\,dy \left[ D_{s,x} (\partial_x \theta)^2 + D_{s,y} (\partial_y \theta)^2 \right].
\end{equation}
We then rescale lengths using $x' = (D_0/D_{s,x})^{1/2} x$ and $y' = (D_0/D_{s,y})^{1/2} y$, where
$D_0$ is any convenient energy scale for normalization, to obtain
\begin{equation}
{\cal F}= {1 \over 2}\ \left( D_{s,x} D_{s,y}  \right)^{1/2} \int dx'\,dy' \left[(\partial_{x'} \theta)^2 + (\partial_{y'} \theta)^2 \right].
\end{equation}
This immediately leads to the generalization of the Nelson-Kosterlitz result in equation~(\ref{eq:aniso-bkt}).
We emphasize that  the reason this seemingly naive argument works is that the line of fixed points below $T_c$ 
are actually described by a Gaussian theory and the BKT $T_c$ is precisely 
when vortex-antivortex unbinding becomes relevant at a Gaussian fixed point.
We thank Steve Kivelson and C. Jayaprakash for very useful conversations related to 
this argument.




\end{document}